 \documentclass[10pt,letter]{iopart}

\usepackage{dcolumn}
\usepackage{graphicx}
\usepackage{epsf}
\usepackage{iopams}
\usepackage{bm}
\usepackage{times}

\def\half{{\textstyle{1\over 2}}}

\def\half{{\textstyle{\frac12}}}

\newcommand{\uu}{{\mathrm u}}
\newcommand{\DD}{{\mathrm D}}
\newcommand{\NN}{{\mathcal N}}

\usepackage{dcolumn}
\newcolumntype{.}{D{x}{}{-1}}

\newcommand{\addrROLLA}{Department of Physics,
Missouri University of Science and Technology,
Rolla, MO 65409-0640, USA}

\begin{document}

\sloppy

\title[Nuclear masses and spectroscopy of Rydberg states]%
{Proposal for the determination of nuclear masses by\\
high-precision spectroscopy of Rydberg states}

\author{B J Wundt and U D Jentschura}
\address{\addrROLLA}

\begin{abstract}
The theoretical treatment of Rydberg states in one-electron ions is facilitated
by the virtual absence of the nuclear-size correction, and fundamental 
constants like the Rydberg constant may be in the reach of planned 
high-precision spectroscopic experiments.
The dominant nuclear effect that shifts transition energies among 
Rydberg states therefore is due to the nuclear mass.
As a consequence, spectroscopic measurements of
Rydberg transitions can be used in order to precisely deduce nuclear masses.
A possible application of this approach to the 
hydrogen and deuterium, and hydrogen-like lithium and carbon is explored in detail.
In order to complete the analysis,
numerical and analytic calculations of the 
quantum electrodynamic (QED) self-energy remainder
function for states with principal quantum number $n=5,\dots,8$
and with angular momentum $\ell=n-1$ and $\ell=n-2$ are described
($j = \ell \pm \half$). 
\end{abstract}

\pacs{06.20.Jr, 12.20.Ds, 31.30.jf\\
{\em Submitted to J.~Phys.~B}}

\maketitle


%
%
\section{Introduction}
\label{intro}

The celebrated high-precision spectroscopic 
experiments of atomic hydrogen~\cite{NiEtAl2000,WeEtAl1995,HuEtAl1998,%
BeEtAl1997,ScEtAl1999,BoEtAl1996,%
BeHiBo1995,HaPi1994,LuPi1986,NeAnUn1979}
typically involve transitions 
among quantum states with low principal quantum numbers,
e.g., the hydrogen $1S$--$2S$ transition.
Unfortunately, the rather large uncertainty of the 
root-mean-square (RMS) charge radius of the proton limits
the precision to which the nominal accuracy of spectroscopic
experiments can be used in order to infer fundamental 
constants: while the most precisely measured transition in 
atomic hydrogen is known to an accuracy of 
1.8 parts in $10^{14}$, the most accurately determined 
fundamental constant (the Rydberg) is known only to an 
accuracy of $6.6$ parts in $10^{12}$ (see Ref.~\cite{MoTaNe2008}).
The subject is not without intricacies: Even the 
correct definition of the RMS charge radius for low-energy processes in atomic systems
is a rather fundamental problem 
which necessitates a careful study of form factors which
are {\em a priori} defined for scattering 
processes~\cite{Pa1996PRA,Pa1999PRA,FrMaSp1997}. 
Indeed, in the 2006 adjustment of the 
fundamental constants~\cite{MoTaNe2008}, the RMS charge radius of the 
proton and deuteron are inferred by fitting the 23 most accurately
measured transitions to QED theory~\cite{JeKoLBMoTa2005}, using the method of least squares.

For the low-lying states canonically used for high-precision
spectroscopy experiments, the dominant nuclear effect is from the 
nuclear size effect on the Dirac energy.
By contrast, the QED theory of transitions among highly excited states
of one-electron ions is simplified because of the 
virtual absence of the problematic nuclear size effect,
and the dominant nuclear proerty that influences the spectrum is the
nuclear mass. This effect is well understood, 
and therefore, the theory of Rydberg transitions among states
with high angular momenta can be formulated very precisely.
In Refs.~\cite{JeMoTaWu2008,JeMoTaWu2009}, Rydberg
states have been proposed in order to avoid these limitations and 
to improve the accuracy of the Rydberg
constant. If we are interested in 
determining the Rydberg constant, then it is important to consider 
hydrogen-like ions with well-known nuclear masses. Conversely,
it becomes possible, using precise QED theory,
to infer the nuclear mass from spectroscopic measurements, if that mass
is not known to sufficient accuracy by other methods. 
That latter aspect forms the basis of the current investigation
and the current proposal which is being investigated in this work.
In selecting suitable atomic Rydberg transitions 
for a conceivable determination of nuclear masses
by this method, it is important to consider frequencies 
accessible to optical frequency combs,
as they constitute one of the most accurate devices available for measurements. 

Consequently, we here investigate the possibility to use high-precision
spectroscopy in order to measure nuclear masses and electron to nucleus 
mass ratios using transitions among Rydberg states
of hydrogen-like ions, with a special emphasis on the example cases of 
${}^1$H, ${}^2$H, ${}^7$Li and ${}^{10}$C.
These masses are of great interest. 
The mass ratios of ${}^1$H and 
${}^2$H are interesting for the frequency comparison of the transitions of 
hydrogen and anti-hydrogen which are being pursued at the 
moment~\cite{GaEtAl2008}, as well as for many other applications of spectroscopy 
in general. ${}^7$Li is
important for the study of solar neutrinos and for determining their 
masses~\cite{Re1967}. There is currently an interesting
discrepancy between the recently obtained mass measurement~\cite{NaEtAl2006} 
and the value in the Atomic Mass Evaluation~\cite{Audi2003337} (AME2003), 
with the more recent value from Ref.~\cite{NaEtAl2006}  being more precise.
Thus, a measurement of the ${}^7$Li mass with a 
completely independent method would seem worthwhile.  
The mass of ${}^{10}$C meanwhile plays a role in super-allowed beta 
decays which are studied in order to measure Cabibbo--Kobayashi--Maskawa 
(CKM) matrix elements~\cite{Blaum2006,HaTo2006}.
The possibility of determining nuclear masses by high-precision
spectroscopy has previously been mentioned in Refs.~\cite{KoEtAl2007,KaEtAl2008}, 
in the context
of molecular spectrosocpy where the theoretical challenges in
providing accurate predictions for mass determination 
appear to be much greater.

In order to determine the mass of the nucleus, we consider essentially two
options. These are based on the elimination of variables in the theoretical
expressions for the transition frequencies and can be summarized as follows.
Namely, up to very high accuracy ($\sim 10^{-14}$), the Rydberg transition
frequencies in one-electron ions are determined by the following parameters: (i)
the nuclear mass and (ii) the Rydberg constant. Our proposed {\em Method I}
applies if at least one isotope of a given charge number has a very well known
mass (of relative accuracy $\sim 10^{-10}$) which can be used as a reference
for the determination of the masses of other isotopes.  We then propose to
measure the same Rydberg transition in two ionic systems
corresponding to two different nuclear isotopes, one of which acts as a
reference, and to solve the system of the two 
equations for two variables, namely the Rydberg constant and
the unknown nuclear mass.  {\em Method II} applies if we suppose that
recent efforts in measuring an improved value of the Rydberg constant (see
Refs.~\cite{JeMoTaWu2008,JeMoTaWu2009,FlEtAl2008}) are crowned with success,
and that an improved value of the Rydberg constant (of relative accuracy
$10^{-13}\dots 10^{-14}$) is available. In that case, the repetition
of a measurement of the same or of a different Rydberg transition
in a one-electron ion of a different 
nuclear isotope or even with a nucleus of a different
charge number can directly lead to 
a determination of the mass of that nucleus
(we essentially solve two observational 
equations corresponding to the two measured frequencies
for the two unknowns, namely, the Rydberg constant and the nuclear mass 
to be determined). {\em Method III} is essentially 
identical to {\em method II} with the only difference being that the 
theoretical expression is solved for 
the electron to nucleus mass ratio instead of the nuclear mass. 

So far, transitions in circular Rydberg states have been measured up to a
relative accuracy of $2.1 \times 10^{-11}$ in an $80\,{\rm K}$ atomic beam of
hydrogen~\cite{DKpc,DVrPhD2001}. The measurement was carried out in the
millimeter region and shows that it is in principle possible to conduct
high-precision measurements on Rydberg states. There is no fundamental
obstacle to improving the experimental accuracy of 
infra-red and near-optical transitions among Rydberg states of hydrogen-like ions
up to the level of $10^{-14} \dots 10^{-15}$.
Using femto-second lasers and frequency combs, this level of 
accuracy has already been attained for the lower-lying
states of hydrogen. Spectroscopy on this level of accuracy has the
potential to assist in the determination of nuclear masses, as we show 
here. In order to carry
out the current study, a supplementary calculation of the self-energy remainder
function is carried out in the range of low nuclear charge numbers $Z=3,6,8$,
for excited states with principal quantum number $n=5,\dots,8$ and orbital
angular momenta $\ell=n-1$ and $\ell=n-2$. 

The outline of this paper implies a discussion of QED theory in
Sec.~\ref{qedcalc}, with a detailed discussion of the proposal for the nuclear
mass determination following in Sec.~\ref{nuclmass}.  The numerical and
analytic methods for the treatment of the self-energy remainder functions are
described in Sec.~\ref{numerics_analytic}.  Numerical examples concerning the
attainable accuracy for the nuclear mass determination are given in
Sec.~\ref{Potential}.  Finally, conclusions are drawn in Sec.~\ref{Conclusions}.

%
%
\section{QED Calculations}
\label{qedcalc}

%
%
\subsection{Status of Theory}
\label{Theo}

In formulating the QED theory of Rydberg transition 
frequencies~\cite{MoTaNe2008,JeMoTaWu2008,JeMoTaWu2009,EiGrSh2001,SaYe1990},
we write the transition frequency between two
Rydberg states of a one-electron atom as
\begin{equation}
\label{nu12}
\nu_{1 \leftrightarrow 2} = \nu_2 - \nu_1 
\end{equation}
for a transition between quantum states $|1 \rangle$ and $|2\rangle$.
The quantities $\nu_i$ ($i=1,2$) are related to the bound-state energies
as $\nu_i = E_i/h$, where $h$ is Planck's constant.
We now discuss the different contributions to the bound-state 
frequencies $\nu_i$ one-by-one, recalling a number of expressions
from Refs.~\cite{MoTaNe2008,JeMoTaWu2008,JeMoTaWu2009,EiGrSh2001,SaYe1990}
which are relevant for the current study. 

The dominant contribution arises from the Dirac
energy, which we convert to a Dirac frequency~$\nu_\DD$. 
We subtract the rest mass and correct for the nuclear mass. For
a state with principal quantum number $n$, total angular momentum $j$, and
angular momentum $\ell$, we have the well-known result
\begin{equation}
\nu_\DD = \frac{R_{\infty} c}{1+r(\NN)} 2
\biggl\{ f(n,j) -1 - \frac{r(\NN) \, \alpha^2}{2\,[1+r(\NN)]^2} \,
[ f(n,j) -1]^2 \biggr\} \,, \label{vdir}
\end{equation}
where $c$ is the speed of light, $R_{\infty} = \alpha^2 m_e c/2 h$ is the
Rydberg constant, $r(\NN) = m_e/m_N(\NN)$ is the mass ratio of the electron to
the nuclear mass, and $\alpha$ is the fine-structure constant. 
In writing $m_N(\NN)$, we use the subscript $N$ in order to denote the 
nuclear mass, and reserve the argument $\NN$ in order to differentiate
a specific nucleus under investigation, e.g., 
$\NN = {}^7$Li or $\NN = {}^{10}$C.
The function $f(n,j)$ is given as
\begin{equation}
f(n, j) = \left[ 1+\frac{(Z \alpha)^2}%
{\left( n-j - \half + \sqrt{(j+ \half)^2 -(Z \alpha)^2} 
\right)^2} \right]^{-\half} \!\! .
\end{equation}
For non-$S$ states, there is a further nuclear-mass dependent 
contribution from the so-called
Barker--Glover (BG) term which originates from the two-body 
Breit Hamiltonian and eliminates the $(n,j)$ degeneracy
of Dirac theory~\cite{BaGl1955},
\begin{eqnarray}
\nu_{\mathrm{BG}} =& \;
\frac{R_{\infty} c}{1+r(\NN)} 
\frac{r(\NN)^2 Z^4 \alpha^2}{n^3 \, [1+r(\NN)]^2} 
\nonumber\\[2ex]
& \; \times \left( \frac{1}{j+\half} - \frac{1}{\ell+\half} \right) \,
(1- \delta_{\ell \, 0} ) \,. \label{vbg}
\end{eqnarray}
The relativistic-recoil (RR) correction changes the frequency of a level with 
$\ell \geq 2$ by~\cite{Er1977,PaGr1995,GoElMiKh1995,JePa1996}
\begin{eqnarray}
\label{vrr} 
\nu_{\mathrm{RR}} =& \; \frac{R_{\infty} c}{1+r(\NN)} 
\frac{2 \, r(\NN) \, Z^5 \alpha^3}{\pi n^3} 
\biggl\{ \frac{1}{[1+r(\NN)]^2} 
\nonumber \\
& \; \quad \times \biggl[ -\frac{8}{3} \ln k_0 (n,\ell) 
- \frac{7}{3} \frac{1}{\ell(\ell+1)(2\ell+1)} \biggr] 
\nonumber \\
& \; + \pi Z \alpha \, [ 1+r(\NN) ]
\nonumber \\
& \; \quad \times \left[3- \frac{\ell(\ell+1)}{n^2} 
\frac{2}{(4l^2-1)(2\ell+3)} \right] \!+ \! \ldots \biggr\}  \,,
\end{eqnarray}
where $\ln k_0$ is the (nonrelativistic) Bethe logarithm
that depends on $n$ and $\ell$. 
The QED radiative corrections for these levels contribute
\begin{eqnarray}
\label{vqed}
\nu_{\mathrm{QED}} =& \; \frac{R_{\infty} c}{1+r(\NN)} 
\frac{2Z^4 \alpha^2}{n^3} 
\biggl\{ - \frac{1}{1+r(\NN)} \frac{a_e}{\kappa(2\ell+1)} 
\nonumber\\
& \; \quad + \frac{1}{[ 1+r(\NN) ]^2} 
\frac{\alpha}{\pi} \biggl[ - \frac{4}{3} \ln k_0 (n,\ell) + \frac{32}{3} 
\nonumber\\
& \times  \frac{3 n^2-\ell(\ell+1)}{n^2} 
\frac{(2\ell-2)!}{(2\ell+3)!} (Z \alpha)^2 
\ln \left( \frac{1+r(\NN)}{(Z \alpha)^2} \right) 
\nonumber\\
& \; \quad + (Z \alpha)^2 \, G(Z \alpha) \biggr] \biggr\} \,,
\end{eqnarray}
where $\kappa = (-1)^{j-\ell+1/2} \, (j + \half)$ is the Dirac angular
quantum number and $a_e$ the electron magnetic moment anomaly.
The semi-analytic expansion of the
self-energy remainder function $G(Z\alpha)$,
which depends on $n$, $\ell$ and $j$, reads 
\begin{eqnarray}
G(Z \alpha) &= A_{60} + (Z \alpha)^2 \, 
\left[ A_{81} \ln (Z \alpha)^{-2} + A_{80} + \ldots \right] \nonumber \\
& \quad + \frac{\alpha}{\pi} B_{60} + \ldots + 
\left( \frac{\alpha}{\pi} \right)^2 C_{60} + \ldots \,. \label{GZ}
\end{eqnarray}
Here, we use the commonly accepted notation for the QED correction terms. The letters
denote the loop order, i.e., $A$ coefficients arise from one-loop diagrams,
$B$ from two-loop, and $C$ from three-loop QED
corrections.  Vacuum polarization is negligible for
Rydberg states with $n \geq 5$, $\ell \geq 3$ and $Z \leq 8$,
on the level of $10^{-15}$ of relative accuracy for the
transitions under study.
So, we can restrict the discussion to the self-energy
contributions to the $A$ coefficients. The first subscript of the $A$ coefficients
denotes the power of $Z \alpha$, while the second denotes the power of the
logarithm $\ln[(Z \alpha)^{-2}]$.  To achieve an appropriate accuracy using
this expansion, the $A_{60}$ coefficients have to be determined. The
higher-order one-photon terms except the vacuum polarization contribution to
$A_{80}$~\cite{WiKr1956}, which is very small, are so far unknown. Calculations
also exist for $B_{60}$ (see Ref.~\cite{Je2006}), however only for states with $\ell
\leq 5$. The three-loop term $C_{60}$ as well as the higher-order two-photon
contributions not listed in Eq.~\eref{GZ} are unknown.

Several results for the $A_{60}$ coefficient of Rydberg states
in the region $n = 9,\ldots, 16$ of principal quantum numbers
have recently been obtained 
(see Refs.~\cite{JeMoTaWu2008,JeMoTaWu2009,WuJe2008}).
The calculation has to be done individually for each one of the 
ionic states, because several logarithmic sums over the 
spectrum of virtual excitations (``relativistic Bethe logarithms'') 
can only be calculated
numerically. In the current investigation, a different 
range of principal quantum number $n = 5, \dots 8$ 
and orbital angular momenta $\ell=n-1$ and $\ell = n-2$ 
will be investigated, because the corresponding transition frequencies
match the most appropriate regions for the application of frequency combs 
to the most interesting isotopes. 
We investigate the one-electron systems of lithium, carbon and oxygen.
The analytic results for $A_{60}$ are
compared to a nonperturbative (in $Z\alpha$) evaluation of the 
self-energy remainder function $G_{\rm SE}(Z\alpha)$,
which corresponds to the entire contribution of higher-order terms
to the one-loop self-energy shift for the ionic states 
under investigation (see Tab.~\ref{tableGSE}). The function $G_{\rm SE}(Z\alpha)$
is equal to the sum of the higher-order $A$ terms, including $A_{60}$,
due to the electron self-energy, as written in Eq.~\eref{GZ}.

For nuclei with nonzero nuclear spin another contribution arises from the interaction of the nuclear spin $I$ with the total angular momentum $j$ of the electron. This leads to the known hyperfine slitting of the energy levels which is given as \cite{JeYe2006}
\begin{eqnarray}
\label{vhfs}
\nu_{\mathrm{hfs}} =& \; \frac{R_{\infty} c}{1+r(\NN)} 
\frac{Z^3 \alpha^2}{n^3} 
\biggl\{ \frac{r(\NN)}{1+r(\NN)} \frac{\kappa}{|\kappa|} 
\nonumber\\
& \; \quad \times \frac{g}{(2 \kappa +1) (\kappa^2 -\textstyle{\frac{1}{4}})}  
\left[ F(F+1) - I(I+1) - j(j+1) \right] 
\nonumber\\
& \times n^3 |\kappa| (2\kappa +1) 
\frac{2 \kappa (\gamma + n-|\kappa|)-N}{N^4 \gamma (4 \gamma^2-1)} \,
\left\{ 1 + 
\frac{\alpha}{\pi} \frac{1}{4 \kappa}\right\} \, ,
\end{eqnarray}
where $g$ is the nuclear $g$ factor, 
$\vec F= \vec I+ \vec j$ the total angular momentum of the
one-electron ion, $\gamma=\sqrt{\kappa^2-(Z \alpha)^2}$ and
$N=\sqrt{(n-|\kappa|)^2+2(n-|\kappa|)\gamma+\kappa^2}$.
The term of relative order $\alpha$ has been obtained 
in Ref.~\cite{BrPa1967}; the generalization to arbitrary 
quantum numbers is found here. 
If necessary, the corrections to the hyperfine splitting
up to relative order $\alpha (Z\alpha)^2$ can be calculated
using a generalization of the approach of 
Ref.~\cite{JeYe2010}.

Having discussed all the required corrections for the 
energy levels of Rydberg states in one-electron ions for the purposes
of our investigation,
we can write down the total frequency of a specific level as the sum
\begin{equation}
\label{nui}
\nu_i = \nu_{\mathrm{D}}+\nu_{\mathrm{BG}} + \nu_{\mathrm{RR}} + 
\nu_{\mathrm{QED}} \,.
\end{equation}
where $i =1,2$ in the spirit of Eq.~\eref{nu12}.
We here assume that the $\nu_{\mathrm{hfs}}$ has been subtracted
from the transition frequency.
if necessary, the hyperfine-fine structure 
mixing terms can be calculated according to the approach 
outlined in Sec.~III of Ref.~\cite{Pa1996PRA}.
However, we note that the state with the highest projection $M_F$
in a given manifold of total electron$+$nuclear angular
momentum $F$ does not receive any corrections
due to hyperfine-structure fine-structure mixing.

%
%
\subsection{Calculation of The Self--Energy Remainder}
\label{numerics_analytic}

The expression~\eref{vqed} for the QED corrections to 
an ionic level includes recoil corrections in the form 
of the electron-to-nuclear mass ratio $r(\NN)$.
For the evaluation of the self-energy correction, it 
is much more practical to use the non-recoil approximation,
which implies an infinitely heavy nucleus and therefore 
the limit $r(\NN) \to 0$. In the non-recoil limit,
the one-photon self-energy shift 
$\Delta E (n \ell_j)$ of 
principal quantum number $n$, orbital angular momentum $\ell$ and 
total angular momentum quantum number $j$ can be written as
\begin{eqnarray}
\label{dE}
& \Delta E (n \ell_j) =
2h R_{\infty} c \, 
\frac{\alpha}{\pi} \frac{Z^4 \alpha^2}{n^3} \biggl\{ A_{40} \\
& \; +  (Z\alpha)^2 \left( A_{61} \ln\left[(Z \alpha)^{-2}\right] + 
G_{\rm SE}(Z\alpha) \right) \biggr\} \,, \;\;\; \ell \geq 2 \,,
\nonumber
\end{eqnarray}
where the first subscript of the $A$ coefficients
denotes the order of $Z \alpha$ while the second denotes the order of
the logarithm $\ln[(Z \alpha)^{-2}]$. 
We have
\begin{equation}
\label{limG}
\lim_{Z\alpha \to 0} G_{\rm SE}(Z\alpha) = A_{60} \,.
\end{equation}
The form of the expansion
in Eq.~\eref{dE} is valid for states with orbital 
angular momentum quantum number $\ell \geq 2$.
The known results for $A_{40}$ and $A_{61}$ for 
Rydberg states read as follows, 
\begin{eqnarray}
\label{A40}
A_{40} &= - \frac{4}{3} \ln k_0 - \frac{1}{2 \kappa (2 \ell +1)} \,, \\[2ex]
\label{A61}
A_{61} &= \frac{32}{3} \frac{3 n^2 - \ell (\ell+1)}{n^2} \frac{(2 \ell-2)!}{(2 \ell+3)!} \,.
\end{eqnarray}
In Eq.~\eref{vqed}, the prefactor $[1+r(\NN)]^{-3}$ as well as 
the $1+r(\NN)$ in the logarithm account for the reduced-mass effects. 
Also, we reemphasize that these results are exclusively due 
to the self-energy effects, and that vacuum-polarization
shifts vanish for states with $\ell \geq 3$ up to  
the order $\alpha (Z\alpha)^{10}$.

The evaluation of $A_{60}$ proceeds via a scale-separation parameter
(overlapping parameter) that is used in order to express the self-energy shift
$\Delta E (n \ell_j)$ in terms of two contributions, one from high-energy
virtual photons and one from low-energy photons whose energy is commensurate
with the atomic binding energy. The overlapping parameter cancels when both
contributions are added, no matter what regularization is used~\cite{WuJe2008}.
Methods from effective field theory can be used in order to simplify the
calculation of the high-energy part
drastically~\cite{CaLe1986,JeCzPa2005,WuJe2008}.  The method of evaluation has
otherwise been described in detail in
Refs.~\cite{JePa1996,JeMoTaWu2008,JeMoTaWu2009}.  For Rydberg states, it
becomes almost indispensable to formulate the logarithmic sums over the virtual
excitations, which contribute to $A_{60}$, in terms of a discretization of the
spectrum on a lattice~\cite{SaOe1989}. In Table~\ref{tableGSE}, we present
results for the states under investigation, which are highly excited Rydberg
states with principal quantum numbers $n=5,\dots,8$ and with the configurations
$\ell = n-1$ and $\ell = n-2$, and $j = \ell \pm \half$ (four states for a given
$n$).

In columns 4--15 of Table~\ref{tableGSE}, we present the 
results of a nonperturbative (in $Z\alpha$) calculation of the 
self-energy remainder function $G_{\rm SE}(Z\alpha)$ for 
one-electron ions with nuclear charge numbers $Z = 3, 6, 8$
(lithium, carbon and oxygen), for the same states that are also relevant 
for the analytic calculation. Note that in view of Eq.~\eref{limG},
the $A_{60}$ coefficient constitutes the limit as $Z\alpha \to 0$ of 
a quantity which is otherwise dependent on the nuclear charge.
Therefore, in Table~\ref{tableGSE}, a specific value of $Z$ has to be given 
along with any $G_{\rm SE}$ (the value of $\alpha$ employed in the 
calculation is $\alpha^{-1} = 137.036$).
For Rydberg states, the dependence of $G_{\rm SE}(Z\alpha)$ on its argument 
is very weak, and the deviation $|G_{\rm SE}(Z\alpha) - A_{60}|$
numerically is less than $10^{-4}$ for all 
states and all nuclear charge numbers under investigation here.
Note that the evaluation of the self-energy remainder function 
for Rydberg states is a highly nontrivial problem~[see \cite{JeMoTa2009}].
In general, when treating Rydberg states both analytically as well as numerically, 
one has to be 
very careful not to terminate the sum over the angular momenta of the 
virtual states too early, because numerically significant contributions are 
due to virtual excitations with high angular momenta
(displaced from the angular momentum of the reference state by one unit).

\begin{table*}[bth]
\begin{flushleft}
\caption{\label{tableGSE}
Values of the $A_{60}$ coefficient obtained with the analytical method
compared to the numerical results for the self-energy remainder function
$G_{\rm SE}(Z\alpha)$ for $n=5, \dots, n=8$ with $\ell=n-1$ and $\ell=n-2$. 
The numbers in parentheses are standard uncertainties in the last figure
and represent the uncertainty of the final numerical integration 
due to the finite number of lattice points (in the case of $A_{60}$),
and the finite number of integration nodes for the 
numerical evaluation of multi-dimensional integrals 
(in the case of $G_{\rm SE}$).}
\resizebox{\textwidth}{!}{%
\begin{tabular}{c@{\hspace{0.2cm}}c@{\hspace{0.2cm}}c@{\hspace{0.2cm}}c@{\hspace{0.5cm}}%
c@{\hspace{0.5cm}}c@{\hspace{0.5cm}}c@{\hspace{0.5cm}}c}
\hline
\hline
\rule[-2mm]{0mm}{8mm}
$n$ & $\ell$ & $2 j$ & $\kappa$ & $A_{60}$ &
$G_{\rm SE}(Z=3)$ & 
$G_{\rm SE}(Z=6)$ & 
$G_{\rm SE}(Z=8)$  \\
\hline
\rule[-2mm]{0mm}{7mm}
5 & 3 & 5 & 3 & $0.002~403~151~41(5)$ & $0.002~41(3)$ & 
 $0.002~43(3)$ & $0.002~45(3)$ \\
\rule[-2mm]{0mm}{7mm}
5 & 3 & 7 & -4 & $0.008~087~015~45(5)$ & $0.008~09(3)$ & 
$0.008~11(3)$ & $0.008~12(3)$  \\
\rule[-2mm]{0mm}{7mm}
5 & 4 & 7 & 4 & $0.000~814~414~71(5)$ & $0.000~82(3)$ & 
$0.000~82(3)$ & $0.000~82(3)$  \\
\rule[-2mm]{0mm}{7mm}
5 & 4 & 9 & -5 & $0.002~412~929~08(5)$ & $0.002~41(3)$ & 
$0.002~41(3)$ & $0.002~42(3)$ \\
\hline
\rule[-2mm]{0mm}{7mm}
6 & 4 & 7 & 4 & $0.000~827~467~81(5)$ & $0.000~83(3)$ & 
$0.000~83(3)$ & $0.000~83(3)$  \\
\rule[-2mm]{0mm}{7mm}
6 & 4 & 9 & -5 & $0.002~748~250~60(5)$ & $0.002~75(3)$ & 
$0.002~75(3)$ & $0.002~75(3)$  \\
\rule[-2mm]{0mm}{7mm}
6 & 5 & 9 & 5 & $0.000~326~676~~27(5)$ & $0.000~32(3)$ & 
$0.000~33(3)$ & $0.000~33(3)$  \\
\rule[-2mm]{0mm}{7mm}
6 & 5 & 11 & -6 & $0.001~008~201~08(5)$ & $0.001~01(3)$ & 
$0.001~01(3)$ & $0.001~01(3)$  \\
\hline
\rule[-2mm]{0mm}{7mm}
7 & 5 & 9 & 5 & $0.000~325~902~82(5)$ & $0.000~32(3)$ & 
$0.000~33(3)$ & $0.000~33(3)$  \\
\rule[-2mm]{0mm}{7mm}
7 & 5 & 11 & -6 & $0.001~141~603~10(5)$ & $0.001~14(3)$ & 
$0.001~14(3)$ & $0.001~14(3)$  \\
\rule[-2mm]{0mm}{7mm}
7 & 6 & 11 & 6 & $0.000~147~439~22(5)$ & $0.000~15(3)$ & 
$0.000~15(3)$ & $0.000~15(3)$  \\
\rule[-2mm]{0mm}{7mm}
7 & 6 & 13 & -7 & $0.000~485~185~97(5)$ & $0.000~48(3)$ & 
$0.000~49(3)$ & $0.000~49(3)$  \\
\hline
\rule[-2mm]{0mm}{7mm}
8 & 6 & 11 & 6 & $0.000~144~496~71(5)$ & $0.000~13(3)$ & 
$0.000~15(3)$ & $0.000~14(3)$  \\
\rule[-2mm]{0mm}{7mm}
8 & 6 & 13 & -7 & $0.000~545~933~41(5)$ & $0.000~54(3)$ & 
$0.000~55(3)$ & $0.000~55(3)$  \\
\rule[-2mm]{0mm}{7mm}
8 & 7 & 13 & 7 & $0.000~072~861~41(5)$ & $0.000~06(3)$ & 
$0.000~07(3)$ & $0.000~07(3)$  \\
\rule[-2mm]{0mm}{7mm}
8 & 7 & 15 & -8 & $0.000~258~766~38(5)$ & $0.000~25(3)$ & 
$0.000~26(3)$ & $0.000~26(3)$  \\
\hline
\hline
\end{tabular}}
\end{flushleft}
\end{table*}

\begin{table}[htb]
\caption{\label{table1}
Fundamental constants and masses used as input parameters for the evaluation of
the theoretical expression and error estimates. In parentheses,
we indicate the standard uncertainty. The masses $m_A(\NN)$ correspond to the 
atomic mass of an atom (including the bound electrons) with nucleus $\NN$.
By contrast, the nuclear mass is denoted as $m_N(\NN)$ 
in this article (it excludes the mass of the bound electrons).}
\begin{indented}
\item[]\begin{tabular}{l@{\hspace{0.2in}}l}
\hline
\hline
\rule[-2mm]{0mm}{6mm}
Constant &
\multicolumn{1}{c}{Value} \\ 
\hline
\rule[-2mm]{0mm}{6mm}
$R_{\infty} c$ & $\phantom{1} 
  3.289 \, 841\, 960\, 361(22) \times 10^{15} \, \mathrm{Hz}$ \\
\rule[-2mm]{0mm}{6mm}
$\alpha$ &       $\phantom{1} 7.297\, 352\, 5376(50) \times 10^{-3}$ \\
\rule[-2mm]{0mm}{6mm}
$a_e$ &          $\phantom{1} 1.159\, 652\, 180\,73(28) \times 10^{-3}$ \\
\rule[-2mm]{0mm}{6mm}
$m_e$ &          $\phantom{1} 5.485\, 799\, 0943(23) \times 10^{-4} \, \uu$ \\
\rule[-2mm]{0mm}{6mm}
$m_A({}^{12}$C) &  
$12.000\, 000\, 000 (0) \, \uu\phantom{5}$ \\
\rule[-2mm]{0mm}{6mm}
$m_A({}^{10}$C) &  
$10.016\, 853\, 2(4) \, \uu\phantom{1}\phantom{5}\phantom{5}$ \\
\rule[-2mm]{0mm}{6mm}
$m_A({}^{6}$Li) & 
$\phantom{1} 6.015\, 122\, 795(16) \, \uu$ \\
\rule[-2mm]{0mm}{6mm}
$m_A({}^{7}$Li) &  
$\phantom{1} 7.016\, 003\, 4256(45) \, \uu$ \\
\rule[-2mm]{0mm}{6mm}
$m_A({}^{1}$H) & 
$\phantom{1} 1.007\, 825\, 032\, 07(10) \, \uu$ \\
\rule[-2mm]{0mm}{6mm}
$m_A({}^{2}$H) &  
$\phantom{1} 2.014\, 101\, 778\, 040(80) \, \uu$ \\
\hline
\hline
\end{tabular}
\end{indented}
\end{table}

\begin{table}[th]
\caption{\label{tableH}
Theoretical predictions for two-photon transition frequencies
in atomic hydrogen and deuterium.
The transition from the initial level $|1\rangle$ with quantum 
numbers $n=9$, $\ell=8$, and $j=15/2$ to the level $|2\rangle$
with quantum numbers $n=16$, $\ell=10$, and $j=19/2$ in
considered. For the upper state, the higher-order self-energy 
coefficient reads $A_{60}(n=16, \ell=10, j=\textstyle{\frac{19}{2}}) = 1.026705(5) \times 10^{-5}$.
The individual contributions are listed in Eq.~\eref{nui}.}
\begin{indented}
\item[]\begin{tabular}{l@{\hspace{0.2in}}l@{\hspace{0.2in}}l}
\hline
\hline
\rule[-2mm]{0mm}{6mm}
Term &
\multicolumn{1}{c}{${}^1{\textrm{H}}$ $\nu$(THz)} &
\multicolumn{1}{c}{${}^2{\textrm{H}}$ $\nu$(THz)} \\ 
\hline
\rule[-2mm]{0mm}{6mm}
$\nu_\DD$ & 
$27.749~282~6987(2)$ & $27.756~833~2542(2)$ \\
\rule[-2mm]{0mm}{6mm}
$\nu_{\mathrm{BG}}$ & 
$\phantom{2}  0.000~000~0000$ & 
$\phantom{2}  0.000~000~0000$ \\
\rule[-2mm]{0mm}{6mm}
$\nu_{\mathrm{RR}}$ & $\phantom{2} 0.000~000~0000$ & 
$\phantom{2} 0.000~000~0000$ \\
\rule[-2mm]{0mm}{6mm}
$\nu_{\mathrm{QED}}$ & $\phantom{2} 0.000~000~0035$ & 
$\phantom{2} 0.000~000~0035$ \\
\rule[-2mm]{0mm}{6mm}
Total & $27.749~282~7022(2)$ & $27.756~833~2577(2)$ \\
\hline
\hline
\end{tabular}
\end{indented}
\end{table}

\begin{table}[th]
\caption{\label{tableLi}
Theoretical predictions for two-photon transition frequencies
in two isotopes of hydrogen-like lithium.
The transition from the initial level $|1\rangle$ with quantum 
numbers $n=9$, $\ell=8$, and $j=15/2$ to the level $|2\rangle$
with quantum numbers $n=16$, $\ell=10$, and $j=19/2$ in
considered. We recall that 
for the upper state, the higher-order self-energy 
coefficient reads $A_{60}(n=16, \ell=10, j=\textstyle{\frac{19}{2}}) = 1.026705(5) \times 10^{-5}$.
Again, the individual contributions are listed in Eq.~\eref{nui}.}
\begin{indented}
\item[]\begin{tabular}{l@{\hspace{0.2in}}l@{\hspace{0.2in}}l}
\hline
\hline
\rule[-2mm]{0mm}{6mm}
Term &
\multicolumn{1}{c}{${}^6{\textrm{Li}}^{2+}$ $\nu$(THz)} &
\multicolumn{1}{c}{$^7{\textrm{Li}}^{2+}$ $\nu$(THz)} \\ 
\hline
\rule[-2mm]{0mm}{6mm}
$\nu_\DD$ & 
$249.857~322~9816(17)$ & $249.860~575~0963(17)$ \\
\rule[-2mm]{0mm}{6mm}
$\nu_{\mathrm{BG}}$ & 
$\phantom{2} \phantom{4}0.000~000~0000$ & 
$\phantom{2} \phantom{4} 0.000~000~0000$ \\
\rule[-2mm]{0mm}{6mm}
$\nu_{\mathrm{RR}}$ & $\phantom{2} \phantom{4}0.000~000~0000$ & 
$\phantom{2} \phantom{4}0.000~000~0000$ \\
\rule[-2mm]{0mm}{6mm}
$\nu_{\mathrm{QED}}$ & $\phantom{2} \phantom{4}0.000~000~2829$ & 
$\phantom{2} \phantom{4} 0.000~000~2829$ \\
\rule[-2mm]{0mm}{6mm}
Total & $249.857~323~2645(17)$ & $249.860~575~3793(17)$ \\
\hline
\hline
\end{tabular}
\end{indented}
\end{table}

\begin{table}[th]
\caption{\label{tableC} Transition frequencies for the transition from level
$|1\rangle$ with quantum numbers $n=13$, $\ell=11$, and $j=21/2$ to level
$|2\rangle$ with quantum numbers $n=17$, $\ell=13$, and $j=25/2$ in two
isotopes of hydrogen-like carbon. The self-energy remainder for the upper state 
is estimated based on a
coefficient of $A_{60}(n=17, \ell=13, j=\textstyle{\frac{25}{2}}) = 3.76900(5) \times 10^{-6}$.}
\begin{indented}
\item[]\begin{tabular}{l@{\hspace{0.2in}}l@{\hspace{0.2in}}l}
\hline
\hline
\rule[-2mm]{0mm}{6mm}
Term &
\multicolumn{1}{c}{$^{12}{\textrm{C}}^{5+}$ $\nu$(THz)} &
\multicolumn{1}{c}{$^{10}{\textrm{C}}^{5+}$ $\nu$(THz)} \\ 
\hline
\rule[-2mm]{0mm}{6mm}
$\nu_\DD$ & 
$ 290.976~045~2425(19)$ & $ 290.973~410~2736(20)$ \\
\rule[-2mm]{0mm}{6mm}
$\nu_{\mathrm{BG}}$ & 
$\phantom{2} \phantom{4} 0.000~000~0000$ & 
$\phantom{2} \phantom{4} 0.000~000~0000$ \\
\rule[-2mm]{0mm}{6mm}
$\nu_{\mathrm{RR}}$ & 
$\phantom{2} \phantom{9} 0.000~000~0001$ & $\phantom{2} \phantom{9}  0.000~000~0001$ \\
\rule[-2mm]{0mm}{6mm}
$\nu_{\mathrm{QED}}$ & 
$\phantom{2} \phantom{9} 0.000~000~6216$ & $\phantom{2} \phantom{9} 0.000~000~6216$ \\
\rule[-2mm]{0mm}{6mm}
Total & $290.976~045~8641(19)$ & $290.973~410~8953(20)$ \\
\hline
\hline
\end{tabular}
\end{indented}
\end{table}

%
%
\subsection{Estimate of Theoretical Uncertainties}

In order to gauge the applicability of our method for determining nuclear
masses, we have to investigate the inevitable theoretical uncertainties that
affect the theoretical predictions of Rydberg transition frequencies in
hydrogen-like ions.  These theoretical uncertainties ultimately determine the
limits of accuracy for the extraction of the nuclear masses.

Two different sources of theoretical uncertainty have to be distinguished.  The
first of these comes from the input parameters necessary to evaluate the
theoretical expression in Sec.~\ref{Theo}. For these input parameters (nuclear
masses and other fundamental constants), we use the values of the fundamental
constants from CODATA 2006~\cite{MoTaNe2008}, the masses from the 2003
Atomic Mass Evaluation~\cite{Audi2003337} (AME2003) and the recent 
measurement for the ${}^{7}$Li mass \cite{NaEtAl2006} which are given in
Table~\ref{table1}. Moreover, the experimental value for $a_e$ 
is used in the theoretical expression~\eref{vqed}; it reads
\begin{equation}
a_e = 1.159 \, 652 \, 180 \, 73(28) \times 10^{-3}
\end{equation}
as obtained with a one-electron cyclotron~\cite{HaFoGa2008}.
We also remember that the 2006 CODATA value of 
the Rydberg constant~(see Table~\ref{table1}) carries a 
relative accuracy of $6.6 \times 10^{-12}$  which currently limits the 
accuracy of theoretical predictions of the transition frequencies. 

Another point concerns the uncertainty
of the reference data for the nuclear masses.
The masses given in Ref.~\cite{NaEtAl2006,Audi2003337} are the atomic masses.
The mass of the bare nucleus needed for 
our calculation can be obtained by
subtracting the masses of the electrons and their binding energies as described
in Ref.~\cite{MoTa2005}. Useful tabulations of ionization energies 
can likewise be found in Ref.~\cite{MoTa2005,CRC}. 

The second source of theoretical uncertainty is due to uncalculated
higher-order terms in the theoretical expressions discussed in Sec.~\ref{Theo}.
The uncertainties of $\nu_{\mathrm{RR}}$ and $\nu_{\mathrm{QED}}$ are
estimated as follows. For the
higher-order recoil terms, we use the magnitude of the last contribution on the
right-hand side of Eq.~(\ref{vrr}) times $(Z \alpha) \ln[(Z \alpha)^{-2}]$. The uncertainty
in $\nu_{\mathrm{QED}}$ is separated into two parts. For the unknown 
$A_{81}$ term in the analytic expansion, we base our estimate on the 
analytic approach and estimate the magnitude of the $A_{81}$ 
term to be equal to $A_{60}$ times $(Z \alpha)^2 \ln[(Z \alpha)^{-2}]$.  
This provides for a uniform estimate which is not restricted to those 
states for which nonperturbative, numerical results for the self-energy 
remainder function are available.
Based on a comparison of the results in states with $\ell \leq 5$,
the magnitude of $B_{60}$ \cite{Je2006} is estimated to be 
equal to the magnitude of $4 \, A_{60}$. This
is used as the uncertainty for this second
term. These estimates are analogous to those used previously in
Ref.~\cite{JeMoTaWu2008}.

We now illustrate, by way of example, the predictive power of QED theory for
Rydberg state transitions in the infra-rad and near optical range, which are
accessible to frequency comb measurements in one-electron ions.  To this end,
we select a two-photon transition in hydrogen, lithium and carbon and add the
theoretical contributions to the transition frequency~\eref{nu12} given in
Sec.~\ref{Theo} in order to obtain a theoretical prediction.  The results are
given in Table~\ref{tableH} for hydrogen and deuterium and in
Table~\ref{tableLi} for two lithium isotopes. In both cases we consider a
two-photon transition from the state $|1 \rangle$ with quantum numbers $n=9$,
$\ell = 8$, $j = 15/2$ to a state $|2\rangle$ with quantum numbers $n=16$,
$\ell = 10$, $j = 19/2$. The 
hydrogen-deuterium isotope shift with
mass numbers $1$ and $2$ as well as for the lithium isotopes with mass numbers $6$
and $7$ is being considered.  The same is done in Table~\ref{tableC} for another
two-photon transition from the state $|1 \rangle$ with quantum numbers $n=13$,
$\ell = 11$, $j = 21/2$ to a state $|2\rangle$ with quantum numbers $n=17$,
$\ell = 13$, $j = 25/2$ in carbon isotopes.  If we wish to extract the Rydberg
constant from a measurement of a two-photon
transition~\cite{JeMoTaWu2008,JeMoTaWu2009}, then we have to consider ions with
very well determined and known nuclear masses.  In our example cases considered
above in Table~\ref{tableH}, the relative accuracy of the electron to nucleus
mass ratio is as follows:
\begin{eqnarray} \frac{\delta r({}^{1}{\textrm{H}})}{r({}^{1}{\textrm{H}})} =&
\; 4.3 \times 10^{-10} \,, \qquad \frac{\delta
r({}^{2}{\textrm{H}})}{r({}^{2}{\textrm{H}})} = 4.2 \times 10^{-10} \,.
\end{eqnarray}
By contrast, the relative accuracy of the nuclear masses for the example cases in
Tables~\ref{tableLi} and~\ref{tableC} is as follows,
\begin{eqnarray} \frac{\delta
m_N({}^{6}{\textrm{Li}})}{m_N({}^{6}{\textrm{Li}})} =& \; 2.7 \times 10^{-9}
\,, \qquad \frac{\delta m_N({}^{7}{\textrm{Li}})}{m_N({}^{7}{\textrm{Li}})} =
6.4 \times 10^{-10} \,, \\[2ex] \frac{\delta
m_N({}^{10}{\textrm{C}})}{m_N({}^{10}{\textrm{C}})} =& \; 4.0 \times 10^{-8}
\,, \qquad \frac{\delta m_N({}^{12}{\textrm{C}})}{m_N({}^{12}{\textrm{C}})} =
1.2 \times 10^{-13} \,.  \end{eqnarray}
Currently, none of the given uncertainties in the mass ratios limit the
final accuracy of the theoretical predictions of Rydberg state transitions (the
theoretical accuracy is limited by the current value of the Rydberg constant,
on a level of about $7 \times 10^{-12}$).  However, in view of a conceivable
improvement of the accuracy of the Rydberg constant in the near future, the
electron to nucleus mass ratios of ${}^{1}$H and ${}^{2}$H as well as the
nuclear masses of ${}^{6}$Li, ${}^{7}$Li and ${}^{10}$C may soon become a
limiting factor.  Specifically, for frequency measurements better than $6.0
\times 10^{-14}$ for ${}^{7}$Li, the nuclear mass accuracy will become
limiting, and for the other nuclei, this effect becomes relevant at a relative
accuracy of about $10^{-13}$ for the frequency measurements.  Conversely,
frequency measurements of better accuracy allow for a better mass determination
using spectroscopy.  The latter aspect will be analyzed in more detail in the
following.

%
%
\section{Nuclear Mass Determination}
\label{nuclmass}

%
%
\subsection{General Paradigms}
\label{massdet}

We intend to explore the applicability of three methods for the determination of
nuclear masses via high-precision spectroscopy of Rydberg transitions in
hydrogen-like ions.  As already outlined in Sec.~\ref{intro}, one of these
methods is essentially based on a very precise measurement of the isotope shift
of a specific Rydberg transition for two isotopes, one of which has a very well
known reference mass. We denote the nucleus with the accurately known reference
mass as $\NN_R$ and its mass as $m_N(\NN_R)$, whereas the other nucleus, whose
mass $m_N(\NN_M)$ is to be determined, is denoted as $\NN_M$.  One
measures two transition frequencies, one in the reference system ($\nu^R_{1
\leftrightarrow 2}$) and one in the isotope whose mass is to be determined
(denoted $\nu^M_{1 \leftrightarrow 2}$).  The system of equations composed of
the two measured transition frequencies 
$\nu^R_{1 \leftrightarrow 2}$ and $\nu^M_{1 \leftrightarrow 2}$ and 
QED theory can then be solved for
the Rydberg constant and for the mass of the unknown isotope. This method is
known as {\em method I} in the current paper.

For {\em method II} and {\em method III}, we assume that various 
efforts of deducing an improved
value for the Rydberg constant~\cite{JeMoTaWu2008,JeMoTaWu2009,FlEtAl2008}
using a Rydberg transition in a hydrogen-like ion with a very well known nuclear
mass are successful.  A second measured transition frequency $\nu^R_{1
\leftrightarrow 2}$ in a system with an inaccurately known nuclear mass
$m_N(\NN_M)$ can directly be compared to its theoretical value. The
nuclear mass $m_N(\NN_M)$ or the electron to nucleus 
mass ratio $r(\NN_M)$ can then be determined.

For the above methods, the following order-of-magnitude estimates are
relevant: We lose about four decimals when converting the frequency measurement
to a measurement of the nuclear mass.  Based on a comparison to high-precision
spectroscopy of lower-lying levels of atomic hydrogen~\cite{FiEtAl2004}, 
we assume that a
reasonable target accuracy for a Rydberg state transition lies in the range of
$10^{-14}$.  A sufficient accuracy of a nuclear reference mass for our purposes
therefore corresponds to the level of $10^{-10}$.  Two Rydberg transition
frequencies of relative accuracy $10^{-14}$, one of which in a reference
system, can be solved for the Rydberg constant (of accuracy $10^{-14}$) and
for the unknown nuclear mass (to be determined with an accuracy of $10^{-10}$,
in the context of {\em method I}).  Alternatively, if the Rydberg constant is
independently known with an accuracy of $10^{-14}$ via a measurement in a
reference system, then any other nuclear mass can be determined with an
accuracy of $10^{-10}$ by a measurement of a different transition in a
different hydrogen-like ion (the 
latter consideration is relevant to {\em method II}).

Let us now cast these order-of-magnitudes estimates into formulas.  We first
have to express the dependence of a transition frequency $\nu_{1
\leftrightarrow 2}$ on the electron to nucleus mass ratio
$r(\NN)=m_e/m_N(\NN)$,
where $m_N(\NN)$ is the mass of the bare nucleus  (without the 
bound electrons).  As can be seen from the theoretical expressions in
Sec.~\ref{Theo}, this dependence has a complicated functional form.  E.g., in
the Dirac value Eq.~(\ref{vdir}), the first term is directly proportional to
the reduced mass $\mu_r=1/(1+r(\NN))$, while the second is proportional to
$r(\NN) \mu_r^3$. However, the dominant and leading
dependence is simply given by the approximate proportionality of the transition
frequency to the reduced mass of the system. We can thus define
a scaled frequency $f_{1 \leftrightarrow 2}$, which is related
to $\nu_{1 \leftrightarrow 2}$ by
\begin{equation} 
\label{scaled}
\nu_{1 \leftrightarrow 2} = 
R_{\infty} c \; \frac{1}{1+r(\NN)} \, f_{1 \leftrightarrow 2} \,.
\end{equation}
The scaled frequency $f_{1 \leftrightarrow 2}$ given by theory still carries a
residual dependence on the mass ratio. For stable/long living nuclei 
equal in mass or heavier than lithium,
known electron to nucleus mass ratios provide
enough accuracy so that the residual dependence of $f_{1 \leftrightarrow 2}$ does
not contribute to the uncertainty on a level required for the nuclear mass determination.

Formulas for {\em method I}: 
We need one isotope with a well-known nuclear mass which we
will use as the reference system. Its transition frequency, mass ratio and
theoretical value are labeled $\nu^R_{1 \leftrightarrow 2}$, 
$r(\NN_R)$, and $f^R_{1 \leftrightarrow 2}$, respectively. Based on this reference,
we want to measure the mass of the nucleus of another isotope. Its
transition frequency, mass ratio and theoretical value will be labeled
$\nu^M_{1 \leftrightarrow 2}$, $r(\NN_M)$, and $f^M_{1 \leftrightarrow 2}$. By
measuring the transition frequencies $\nu^R_{1 \leftrightarrow 2}$ and
$\nu^M_{1 \leftrightarrow 2}$ in both isotopes, we get the system of equations
\begin{eqnarray}
\nu^R_{1 \leftrightarrow 2} = 
R_{\infty} c \; \frac{1}{1+r(\NN_R)} \, f^R_{1 \leftrightarrow 2} \,, \\
\nu^M_{1 \leftrightarrow 2} = 
R_{\infty} c \; \frac{1}{1+r(\NN_M)} \, f^M_{1 \leftrightarrow 2} \,.\\
\nonumber
\end{eqnarray}
We cancel the Rydberg constant and obtain
\begin{eqnarray}
\frac{\nu^R_{1 \leftrightarrow 2}}{\nu^M_{1 \leftrightarrow 2}} &= \;
\frac{f^R_{1 \leftrightarrow 2}}{f^M_{1 \leftrightarrow 2}} \,
\frac{1+r(\NN_M)}{1+r(\NN_R)} 
\nonumber\\[2ex]
&=  \;
\frac{m_N(\NN_R) 
f^R_{1 \leftrightarrow 2}}{m_N(\NN_M) 
f^M_{1 \leftrightarrow 2}} \;
\frac{m_N(\NN_M)+m_e}{m_N(\NN_R)+m_e} \,.
\end{eqnarray}
Solving for the nuclear mass $m_N(N_M)$, we obtain
\begin{eqnarray}\label{meth1}
m_N(\NN_M) =& \; m_N(\NN_R) \\[2ex]
& \; \times m_e \left[ 
\frac{\nu^R_{1 \leftrightarrow 2} \, f^M_{1 \leftrightarrow 2}}%
{\nu^M_{1 \leftrightarrow 2} \, f^R_{1 \leftrightarrow 2}} \, 
(m_N(\NN_R)+m_e) -  m_N(\NN_R) \right]^{-1} \,.
\nonumber
\end{eqnarray}
This allows us to determine the nuclear mass of one isotope 
$m_N(\NN_M)$ from a measurement
of a transition frequency $\nu^M_{1 \leftrightarrow 2}$
in this isotope, and a reference transition frequency
$\nu^R_{1 \leftrightarrow 2}$ in an isotope with nuclear
mass $m_N(\NN_R)$.

Formulas for {\em method II}: We have already mentioned a joint theoretical and
experimental project with the National Institute of Standards and Technology
(NIST), whose aim is to deduce an improved value for the Rydberg constant from
transitions in Rydberg states~\cite{JeMoTaWu2008,JeMoTaWu2009}. Furthermore,
we also mention a project at the National Physics Laboratory (NPL) in the United
Kingdom where the $2S$--$8D$ transition in hydrogen is intended to be 
used in order to improve the
accuracy of the Rydberg constant \cite{FlEtAl2008}.  Let us assume that one of
these efforts is met with success and the uncertainty of the Rydberg constant
can be reduced significantly (by at least an order of magnitude
as compared to the 2006 CODATA value~\cite{MoTaNe2008}). 

This would allow to open up another possible way of measuring masses which is
especially interesting for isotope systems where no mass is known well enough
to serve as a reference mass. Namely, provided the Rydberg constant
can be determined to good accuracy in a different ionic system,
the nuclear mass $m_N(\NN_M)$ can be obtained by solving 
Eq.~\eref{scaled} yielding
\begin{equation}
\label{meth2} m_N(\NN_M) =  m_e \left( \frac{f_{1 \leftrightarrow
2}^M R_{\infty}c}{\nu^M_{1 \leftrightarrow 2}} -1 \right)^{-1} \,.
\end{equation}
Numerical loss is incurred because
\begin{equation} 
\label{loss}
\frac{f^M_{1 \leftrightarrow 2} \, R_{\infty}c}%
{\nu^M_{1 \leftrightarrow 2}} - 1 = r(\NN_M) \,,
\end{equation}
where $r(\NN_M)$ is rather small ($\approx 10^{-4} \dots 10^{-5}$ in 
typical cases), whereas the two terms on the left hand side are of order unity.

This equation also allows for a determination of the electron to 
nucleus mass ratios which we will denote as {\em method III} in 
the following. In this case, we just use
\begin{equation} 
\label{meth3}
r(\NN_M) = \frac{f^M_{1 \leftrightarrow 2} \, R_{\infty}c}%
{\nu^M_{1 \leftrightarrow 2}} - 1 \,.
\end{equation}
in order to determine $r(\NN_M)$, again with a loss in numerical
significance of above four decimals.

\begin{table}[htb]
\caption{\label{tablem1} We list the sources for the relative uncertainties for
the determination of $\delta m_N(\NN_M)/m_N(\NN_M)$ We explore {\em method I}
for the determination of a nuclear mass (see text). The reference nucleus is
$\NN_R= {}^{12}$C, and the nuclear mass of $\NN_M= {}^{10}$C is to be measured.
The transition is from state $\left| 1 \right> \leftrightarrow \left| 2
\right>$, where $\left| 1 \right>$ has quantum numbers $n=13$, $\ell=11$ and
$j=21/2$, whereas $\left| 2 \right>$ has quantum numbers $n=17$, $\ell=13$ and
$j=25/2$. The contributions to the relative uncertainty $\delta
m_N(\NN_M)/m_N(\NN_M)$ above the horizontal line are evaluated in terms of the
2006~CODATA recommended values of the fundamental constants and do not
influence the determination of $m_N(\NN_M)$ on the level of one part in
$10^{-10}$.  The contributions below the horizontal line are due to the assumed
experimental (spectroscopic) accuracy of the transitions.}
\begin{indented}
\item[]\begin{tabular}{l@{\hspace{0.2in}}c@{\hspace{0.1in}}c}
\hline
\hline
\rule[-2mm]{0mm}{8mm}
Source &
$\displaystyle \frac{\delta m_N(\NN_M)}{m_N(\NN_M)}$ \\[2ex]
\hline
\rule[-2mm]{0mm}{8mm}
$\displaystyle \frac{1}{m_N(\NN_M)} 
\frac{\partial m_N(\NN_M)}{\partial m_N(\NN_R)} \,
\delta m_N(\NN_R)$ & $9.6 \times 10^{-14}$ \\[3ex]
\rule[-2mm]{0mm}{6mm}
$\displaystyle \frac{1}{m_N(\NN_M)} 
\frac{\partial m_N(\NN_M)}{\partial r(\NN_M)}  \,
\delta r(\NN_M)$ & $ 1.8 \times 10^{-13}$ \\[3ex]
\rule[-2mm]{0mm}{6mm}
$\displaystyle \frac{1}{m_N(\NN_M)} 
\frac{\partial m_N(\NN_M)}{\partial r(\NN_R)}  \,
\delta r(\NN_R)$ & $ 1.6 \times 10^{-15}$ \\[3ex]
\rule[-2mm]{0mm}{6mm}
$\displaystyle \frac{1}{m_N(\NN_M)} 
\frac{\partial m_N(\NN_M)}{\partial \alpha}  \,
\delta \alpha$ & $ 1.0 \times 10^{-15}$ \\[3ex]
\rule[-2mm]{0mm}{6mm}
$\displaystyle \frac{1}{m_N(\NN_M)} 
\frac{\partial m_N(\NN_M)}{\partial a_e}  \,
\delta a_e$ & $ 8.8 \times 10^{-20}$  \\[3ex]
\rule[-2mm]{0mm}{6mm}
$\displaystyle \frac{1}{m_N(\NN_M)} 
\frac{\partial m_N(\NN_M)}{\partial m_e}  \,
\delta m_e$ & $6.9 \times 10^{-11}$ \\[3ex]
\rule[-2mm]{0mm}{6mm}
$\displaystyle \delta [f^M_{1\leftrightarrow 2}/f^R_{1\leftrightarrow 2}]$ 
 & $1.6 \times 10^{-11}$ \\[1ex]
\hline
\rule[-2mm]{0mm}{8mm}
$\displaystyle \frac{1}{m_N(\NN_M)} 
\frac{\partial m_N(\NN_M)}{\partial \nu^R_{1\leftrightarrow 2}} \,
\delta \nu^R_{1\leftrightarrow 2}$ & 
$\displaystyle 1.8 \times 10^{4} 
\left(\frac{\delta \nu^R_{1\leftrightarrow 2}}{\nu^R_{1\leftrightarrow 2}} \right)$ 
\\[3ex]
\rule[-2mm]{0mm}{6mm}
$\displaystyle \frac{1}{m_N(\NN_M)} 
\frac{\partial m_N(\NN_M)}{\partial \nu^M_{1\leftrightarrow 2}} \,
\delta \nu^M_{1\leftrightarrow 2}$ & 
$\displaystyle 1.8 \times 10^{4} 
\left(\frac{\delta \nu^M_{1\leftrightarrow 2}}{ \nu^M_{1\leftrightarrow 2}} \right)$ 
\\[3ex]
\hline
\hline
\end{tabular}
\end{indented}
\end{table}
\begin{table}[htb]
\caption{\label{tablem2}
We explore the application of {\em method II} (see text)
for the determination of the mass of the $\NN_M= {}^{7}$Li nucleus.
The transition is $\left| 1 \right> \leftrightarrow \left| 2 \right>$
where $\left| 1 \right>$ is the state with quantum numbers 
$n=9$, $\ell=8$, and $j=15/2$, and $\left| 2 \right>$ has quantum numbers
$n=16$, $\ell=10$ and $j=19/2$.
Contributions because of theoretical input data to the relative 
uncertainty $\delta m_N(\NN_M)/m_N(\NN_M)$ of the nuclear mass
are given above the horizontal line, 
whereas contributions due to the assumed spectroscopic measurements 
are given below the horizontal line.}
\begin{indented}
\item[]\begin{tabular}{l@{\hspace{0.2in}}c@{\hspace{0.2in}}c}
\hline
\hline
\rule[-2mm]{0mm}{8mm}
Source &
$\displaystyle \frac{\delta m_N(\NN_M)}{m_N(\NN_M)}$ \\[2ex]
\hline
\rule[-2mm]{0mm}{8mm}
$\displaystyle \frac{1}{m_N(\NN_M)} 
\frac{\partial m_N(\NN_M)}{\partial r(\NN_M)} \,
\delta r(\NN_M)$ & $ 1.5 \times 10^{-15}$ \\[3ex]
\rule[-2mm]{0mm}{6mm}
$\displaystyle \frac{1}{m_N(\NN_M)} 
\frac{\partial m_N(\NN_M)}{\partial \alpha}  \,
\delta \alpha$ & $ 4.4 \times 10^{-11}$ \\[3ex]
\rule[-2mm]{0mm}{6mm}
$\displaystyle \frac{1}{m_N(\NN_M)} 
\frac{\partial m_N(\NN_M)}{\partial a_e}  \,
\delta a_e$ & $ 3.6 \times 10^{-15}$  \\[3ex]
\rule[-2mm]{0mm}{6mm}
$\displaystyle \frac{1}{m_N(\NN_M)} 
\frac{\partial m_N(\NN_M)}{\partial m_e}  \,
\delta m_e$ & $4.2 \times 10^{-10}$ \\[3ex]
\rule[-2mm]{0mm}{6mm}
$\displaystyle \delta f^M_{1\leftrightarrow 2}$ & $1.1 \times 10^{-11}$ \\[1ex]
\hline
\rule[-2mm]{0mm}{8mm}
$\displaystyle \frac{1}{m_N(\NN_M)} 
\frac{\partial m_N(\NN_M)}{\partial \nu^M_{1\leftrightarrow 2}} \,
\delta \nu^M_{1\leftrightarrow 2}$ & 
$\displaystyle 1.3 \times 10^{4} \,
\left(\frac{\delta \nu^M_{1\leftrightarrow 2}}{\nu^M_{1\leftrightarrow 2}} \right)$ 
\\[3ex]
\rule[-2mm]{0mm}{6mm}
$\displaystyle \frac{1}{m_N(\NN_M)} 
\frac{\partial m_N(\NN_M)}{\partial R_{\infty} c} \,
\delta R_{\infty} c$ & 
$\displaystyle 1.3 \times 10^{4} \,
\left(\frac{\delta R_{\infty} c}{R_{\infty} c} \right)$ 
\\[2ex]
\hline
\hline
\end{tabular}
\end{indented}
\end{table}
\begin{table}[htb]
\caption{\label{tablem3D}
We explore the application of {\em method III} (see text)
for the determination of the electron to nucleus mass ratio 
for $\NN_M= {}^{2}$H.
The transition is $\left| 1 \right> \leftrightarrow \left| 2 \right>$
where $\left| 1 \right>$ is the state with quantum numbers 
$n=9$, $\ell=8$, and $j=15/2$, and $\left| 2 \right>$ has quantum numbers
$n=16$, $\ell=10$ and $j=19/2$. The contributions to the relative 
uncertainty $\delta r(\NN_M)/r(\NN_M)$ of the electron to 
deuteron mass ratio due to the 2006 CODATA 
values for the fundamental constants and masses 
required for the evaluation of the theoretical 
expressions are given above the horizontal
line. Contributions due to the assumed spectroscopic measurements 
are given below the horizontal line.}
\begin{indented}
\item[]\begin{tabular}{l@{\hspace{0.2in}}c@{\hspace{0.2in}}c}
\hline
\hline
\rule[-2mm]{0mm}{8mm}
Source &
$\displaystyle \frac{\delta r(\NN_M)}{r(\NN_M)}$ \\[2ex]
\hline
\rule[-2mm]{0mm}{8mm}
$\displaystyle \frac{1}{r(\NN_M)} 
\frac{\partial r(\NN_M)}{\partial r(\NN_M)} \,
\delta r(\NN_M)$ & $ 9.1 \times 10^{-17}$ \\[3ex]
\rule[-2mm]{0mm}{6mm}
$\displaystyle \frac{1}{r(\NN_M)} 
\frac{\partial r(\NN_M)}{\partial \alpha}  \,
\delta \alpha$ & $ 1.4 \times 10^{-12}$ \\[3ex]
\rule[-2mm]{0mm}{6mm}
$\displaystyle \frac{1}{r(\NN_M)} 
\frac{\partial r(\NN_M)}{\partial a_e}  \,
\delta a_e$ & $ 1.2 \times 10^{-16}$  \\[3ex]
\rule[-2mm]{0mm}{6mm}
$\displaystyle \delta f^M_{1\leftrightarrow 2}$ & $6.8 \times 10^{-33}$ \\[1ex]
\hline
\rule[-2mm]{0mm}{8mm}
$\displaystyle \frac{1}{r(\NN_M)} 
\frac{\partial r(\NN_M)}{\partial \nu^M_{1\leftrightarrow 2}} \,
\delta \nu^M_{1\leftrightarrow 2}$ & 
$\displaystyle 3.7 \times 10^{3} \,
\left(\frac{\delta \nu^M_{1\leftrightarrow 2}}{\nu^M_{1\leftrightarrow 2}} \right)$ 
\\[3ex]
\rule[-2mm]{0mm}{6mm}
$\displaystyle \frac{1}{r(\NN_M)} 
\frac{\partial r(\NN_M)}{\partial R_{\infty} c} \,
\delta R_{\infty} c$ & 
$\displaystyle 3.7 \times 10^{3} \,
\left(\frac{\delta R_{\infty} c}{R_{\infty} c} \right)$ 
\\[2ex]
\hline
\hline
\end{tabular}
\end{indented}
\end{table}
\begin{table}[htb]
\caption{\label{tablem3H}
We explore the application of {\em method III} (see text)
for the determination of the electron to nucleus mass ratio 
for $\NN_M= {}^{1}$H.
The transition is $\left| 1 \right> \leftrightarrow \left| 2 \right>$
where $\left| 1 \right>$ is the state with quantum numbers 
$n=9$, $\ell=8$, and $j=15/2$, and $\left| 2 \right>$ has quantum numbers
$n=16$, $\ell=10$ and $j=19/2$.
Again, contributions to the relative uncertainty $\delta r(\NN_M)/r(\NN_M)$ 
of the mass ratio are separated into those caused
by theoretical input data which are given above the horizontal
line, and contributions due to the assumed spectroscopic measurements 
are given below the horizontal line.}
\begin{indented}
\item[]\begin{tabular}{l@{\hspace{0.2in}}c@{\hspace{0.2in}}c}
\hline
\hline
\rule[-2mm]{0mm}{8mm}
Source &
$\displaystyle \frac{\delta r(\NN_M)}{r(\NN_M)}$ \\[2ex]
\hline
\rule[-2mm]{0mm}{8mm}
$\displaystyle \frac{1}{r(\NN_M)} 
\frac{\partial r(\NN_M)}{\partial r(\NN_M)} \,
\delta r(\NN_M)$ & $ 9.3 \times 10^{-17}$ \\[3ex]
\rule[-2mm]{0mm}{6mm}
$\displaystyle \frac{1}{r(\NN_M)} 
\frac{\partial r(\NN_M)}{\partial \alpha}  \,
\delta \alpha$ & $ 7.0 \times 10^{-13}$ \\[3ex]
\rule[-2mm]{0mm}{6mm}
$\displaystyle \frac{1}{r(\NN_M)} 
\frac{\partial r(\NN_M)}{\partial a_e}  \,
\delta a_e$ & $ 5.8 \times 10^{-17}$  \\[3ex]
\rule[-2mm]{0mm}{6mm}
$\displaystyle \delta f^M_{1\leftrightarrow 2}$ & $1.3 \times 10^{-32}$ \\[1ex]
\hline
\rule[-2mm]{0mm}{8mm}
$\displaystyle \frac{1}{r(\NN_M)} 
\frac{\partial r(\NN_M)}{\partial \nu^M_{1\leftrightarrow 2}} \,
\delta \nu^M_{1\leftrightarrow 2}$ & 
$\displaystyle 1.8 \times 10^{3} \,
\left(\frac{\delta \nu^M_{1\leftrightarrow 2}}{\nu^M_{1\leftrightarrow 2}} \right)$ 
\\[3ex]
\rule[-2mm]{0mm}{6mm}
$\displaystyle \frac{1}{r(\NN_M)} 
\frac{\partial r(\NN_M)}{\partial R_{\infty} c} \,
\delta R_{\infty} c$ & 
$\displaystyle 1.8 \times 10^{3} \,
\left(\frac{\delta R_{\infty} c}{R_{\infty} c} \right)$ 
\\[2ex]
\hline
\hline
\end{tabular}
\end{indented}
\end{table}
%

%
%
\subsection{Potential of the Method}
\label{Potential}

We now illustrate the two discussed methods for the determination
of nuclear masses from Rydberg transition spectroscopy in 
hydrogen-like ions. To this end, we keep the attained experimental 
accuracy as a variable and show to which accuracy 
nuclear masses can be determined for a given assumed
spectroscopic accuracy. We concentrate on transitions
in isotopes of atomic hydrogen, lithium and carbon
that lie in the infra-red frequency range, which is 
an ideal application range for frequency combs. 

For this illustration, we use the atomic mass unit which is also the unit
generally used in measurements in Penning traps. There the masses are
determined by a comparison of cyclotron frequencies of different ions. Most
atomic masses and especially the mass of the electron which is rather important
for our methods are in fact known more precisely in the atomic mass system.
Even though there are at the moment numerous efforts to use the atomic mass unit
to define the SI unit kilogram, so far the conversion factor between the two
units still carries a rather large relative uncertainty of $5.0 \times
10^{-8}$. In the determination of mass ratios (close to unity) or mass differences,
the uncertainty of the conversion factors become irrelevant, and these are
the quantities that one needs for applications to fundamental physics.
We can thus safely work in atomic mass units.

We first discuss an example application for {\em method I}, which as 
we recall implies the determination of 
Rydberg constant and nuclear mass from an accurate measurement 
of two transitions in hydrogen-like ions of different 
isotopes. The reference system is taken as the
one-electron ion of $^{12}$C. We recall that the $^{12}$C atom serves as the
definition of the atomic mass unit (u) which is defined by the relation
$m_A(^{12}\textrm{C})=12\,\uu$.  The nuclear mass $m_N(^{12}\textrm{C})$ is
obtained from $m_A(^{12}\textrm{C})$ by the subtraction of the electron rest
masses and the binding energies. The latter carries a theoretical uncertainty which
has been discussed in~\cite{MoTa2005}; yet $m_N(^{12}\textrm{C})$ still represents a very accurate
nuclear mass standard. Even taking into account this
uncertainty, the final relative uncertainty of the nuclear mass 
is $1.2 \times 10^{-13}$ which is still two orders of magnitude more accurate 
than any measured nuclear mass.  The isotope whose nuclear mass we want to determine is assumed to
be the one-electron ion of $^{10}$C. In both isotopes, the same transition is
employed for the mass determination. The currently best value for the atomic
mass of $^{10}$C is given in Table~\ref{table1} and has a relative uncertainty
of $4 \times 10^{-8}$.  Although this mass determination to 4 parts in $10^8$
is very accurate on an absolute scale, we should remember that there appears to
be room for improvement.  For comparison, one of the most precise mass
measurements has recently been conducted for $^{16}$O with a relative precision
of $1.1 \times 10^{-11}$ (see Ref.~\cite{VDyEtAl2006}). 

In terms of our notation for the ``master equation'' Eq.~(\ref{meth1}) for {\em
method I}, we have $\NN_R= {}^{12}$C and $\NN_M= {}^{10}$C. 
We specify $\left| 1 \right> \leftrightarrow \left| 2 \right>$
as the same transition whose frequency (based on the current values of the fundamental
constants~\cite{MoTaNe2008}) has already been evaluated in Table~\ref{tableC}.
Above the horizontal line of Table~\ref{tablem1}, we list the contributions to
the relative uncertainty $\delta m_N(\NN_M)/m_N(\NN_M)$ of the measured mass of
the $\NN_M= {}^{10}$C due to the reference nucleus mass, due to the
electron-to-nuclear mass ratios  $r(\NN_M)$ and
$r(\NN_R)$, due to the 2006~CODATA value of the fine-structure constant, due to
the experimental uncertainty \cite{HaFoGa2008} of the electron magnetic moment anomaly $a_e$ which
enters the theoretical expression in Eq.~\eref{vqed},
and due to the higher-order QED terms which
enter the theoretical expressions for the scaled transition frequencies $f^M_{1
\leftrightarrow 2}$ and $f^R_{1 \leftrightarrow 2}$. 
Our chosen transition has $j=\ell-\half$ for both states involved.
In general, states with $j=\ell-\half$ allow 
for QED predictions with slightly smaller uncertainties than those with
$j=\ell + \half$. This is because states with $j=\ell-\half$ in general 
have numerically smaller coefficients in the semi-analytic representation
of radiative, radiative-recoil and recoil corrections as compared to states
with $j=\ell + \half$. This trend is expected to 
continue for the uncalculated higher-order terms.
Our procedure for estimating the higher-order
terms, which expresses the estimates in terms of multiples of known
coefficients, therefore leads to slightly smaller theoretical
uncertainties for the mentioned transitions as compared to a case with 
$j=\ell + \half$. 

If we assume that the experimental accuracy for the 
spectroscopic measurements of $\delta \nu^R_{1 \leftrightarrow 2}$ and 
$\delta \nu^M_{1 \leftrightarrow 2}$ is of the order of $10^{-14}$, 
then the by far dominant
contribution to the uncertainty of such a mass determination is 
caused by the uncertainty in the frequency measurements.
Their contribution to $\delta m_N(\NN_M)/m_N(\NN_M)$ is roughly four
orders of magnitude larger than 
$\delta \nu^R_{1 \leftrightarrow 2}/\nu^R_{1 \leftrightarrow 2}$ and
$\delta \nu^M_{1 \leftrightarrow 2}/\nu^M_{1 \leftrightarrow 2}$,
as given below
the horizontal line in Table~\ref{tablem1}. 
Finally, to give a numerical example,
we conclude that 
assuming a relative uncertainty in the frequency measurements
of the order of $1.4 \times 10^{-12}$ could be reached, the 
nuclear mass of $^{10}$C
could be determined with a relative uncertainty of $3.6 \times 10^{-8}$ 
on a level comparable to the 2006 CODATA value.
We hereby add all contributing uncertainties quadratically.
For completeness, we note that 
even if QED theory carried no uncertainty, and even if 
our spectroscopic accuracy were perfect, the maximum accuracy reachable
with {\em method I} would still be limited by the uncertainty of the reference mass. 
Therefore having a reference mass with a high accuracy such as $^{12}$C is recommendable 
in order to minimize one particularly problematic  source 
of uncertainty.

We now discuss the application of {\em method II},
which implies a measurement of a single, specific 
transition in a hydrogen-like ion, with the Rydberg constant
having been determined in a different system.
The ``master equation'' for this method is
Eq.~(\ref{meth2}). Here, the example we study 
is $\NN_M= {}^{7}$Li, with the transition given in 
Table~\ref{tableLi}. The current best value 
\cite{NaEtAl2006} for the mass of $^{7}$Li 
has a relative uncertainty of $6.4 \times 10^{-10}$.
The difference, however, to the previously accepted 
value for the mass of $^{7}$Li \cite{Audi2003337} is 
14 standard deviations or $1.1 \times 10^{-6} \uu$.
Another measurement with an independent method 
with a comparable accuracy to \cite{NaEtAl2006} 
would therefore be of great interest.
In Table~\ref{tablem2}, we list the different sources 
of theoretical uncertainty for the mass determination.
As evident from the entry in the fifth row of Table~\ref{tablem2},
the currently accepted value of the electron mass
also is significant for the mass determination of the lithium isotope,
though not the primary limiting factor. It is 
generally assumed that an improved value of the 
electron mass could be within reach of an improved measurement 
of a bound-electron $g$ factor in a low--$Z$ hydrogen-like 
ion~\cite{VeEtAl2004,VoEtAl2005,WeEtAl2006,QuNiJe2008,QuEtAl2008}.

If it would be possible to reach an accuracy of the frequency 
measurements in the 
range of $2 \times 10^{-14}$ for both the Rydberg determination
frequency as well as the 
Lithium ion frequency (in the setting discussed above), then the nuclear mass of
$^{7}$Li could be determined with a relative uncertainty of $5.6 \times 10^{-10}$.
Thereby, a similar accuracy as in the measurement of \cite{NaEtAl2006} could 
be reached.

In order to examine whether the required precision in the measurement 
can be reached for the discussed transitions, the ratio of the transition 
energy $E$ to the decay width of the line $\Gamma$ has to be 
considered. In \cite{JeMoTaWu2008}, based on a calculation for 
the decay width for Rydberg states carried out in \cite{JeEtAl2005}, 
this $Q$ factor has been 
evaluated for transitions of near circular Rydberg states from $n$ to 
$n-1$ as
\begin{equation}
Q= \frac{E_n-E_{n-1}}{\Gamma_n+\Gamma_{n-1}} = 
\frac{3 n^2}{4 \alpha (Z \alpha)^2} \,.
\end{equation}
For the transitions considered here a complete calculation of the 
decay width following the description in \cite{BeSa1957} leads 
to the results for the $Q$ factor
\begin{eqnarray}
Q({\textrm{Li}})[n=12,\ell=10 \rightarrow n=9, \ell = 8] &= \; 8.0 \times 10^{7} \,, \\[2ex]
Q({\textrm{H}})[n=12,\ell=10 \rightarrow n=9, \ell = 8] &= \; 7.2 \times 10^{8} \,, \\[2ex]
Q({\textrm{C}})[n=17,\ell=13 \rightarrow n=13, \ell = 11] &= \; 2.8 \times 10^{7} \,.
\end{eqnarray}
The comparison of the $Q$ factors for hydrogen and lithium show 
its dependence on the nuclear charge number $Z$. 
Measurements in the past have been able to determine the 
energy of a transition within $10^{-4}$ of the width of the 
line \cite{HaPi1994,LuPi1981}. It would be 
required to enhance this to about $10^{-6}$ to reach the required 
accuracy for a mass determination in Li and to $3 \times 10^{4}$ for 
a mass determination in C. The attempt at 
NIST \cite{JeMoTaWu2008,JeMoTaWu2009} to measure transitions among Rydberg states up to an 
accuracy of $10^{-14}$ might requires one to develop techniques 
which enhance the resolution of the lines to the indicated  values. 

Due to the higher $Q$ values for hydrogen, the uncertainty 
with which the transition frequencies can be determine would 
be about one order of magnitude smaller. Moreover, the electron 
to proton or deuteron mass ratios are much larger compared to mass ratios for 
nuclei with higher $Z$ considered so far. Because 
the numerical loss is directly related to the mass 
ratio [see Eq.~(\ref{loss})], another source of uncertainty is reduced.

This leads us to the discussion of the application of {\em method III}.
Here, we investigate the two cases $\NN_M={}^1$H and 
$\NN_M={}^2$H using the transitions in Tab.~\ref{tableH} with 
the master equation (\ref{meth3})
of {\em method III}. The theoretical sources 
of uncertainty are given in Table~\ref{tablem3D}. 
Based on Refs.~\cite{HaPi1994,LuPi1981}, we
assume that the line can be
split to at least one part in $10^{4}$.
Together with the $Q$ factor this leads to a conservative
estimate of the relative accuracy of the frequency measurement 
of $1.4 \times 10^{-13}$, which we will also use for the assumed 
relative uncertainty of the Rydberg constant. These uncertainties 
would allow to determine the electron to deuteron mass ratio 
with a relative uncertainty of $7.3 \times 10^{-10}$;
this is comparable to the 2006 CODATA mass ratio which has
a relative uncertainty 
of $4.2 \times 10^{-10}$ \cite{MoTaNe2008}. Every improvement in the 
resolution of the line 
would allow to increase the accuracy with which the 
mass ratio can be determined. For $\NN_M={}^1$H, the theoretical 
sources of uncertainty are listed in Table~\ref{tablem3H}. We use 
the same conservative estimate as for ${}^2$H regarding 
the uncertainty of the 
measurement of the transition frequency and the Rydberg constant. 
Even so, the electron to proton mass ratio could still be determined 
with a relative uncertainty of $3.6 \times 10^{-10}$;
this would be slightly better 
than the current best value which has a relative uncertainty 
of $4.3 \times 10^{-10}$ \cite{Audi2003337}.

%
%

\section{CONCLUSIONS}

\label{Conclusions}

The primary goal of the current work has been to show that high-precision
spectroscopy of Rydberg states in hydrogen-like ions has a far greater potential
for advances in fundamental physics than just the determination of the Rydberg
constant and can be used for the determination of nuclear masses as well.  This
finding enhances the likely impact of a successful implementation of
high-precision spectroscopy of highly excited Rydberg states of hydrogen-like
ions with a low and medium nuclear charge number.  In general, we can say that
the basic idea of the current paper is to use a lightly bound electron (in a
Rydberg state) to probe the nucleus (a ``Rydberg electron trap'' for the
nucleus).  The interaction of the nucleus to the loosely bound electron could
be measured with high precision.  Because this interaction is well understood,
it can be used to determine the mass of the nucleus. Hyperfine effects caused 
by the interaction of the nuclear spin with the total angular momentum of 
the electron are briefly considered here in Eq.~(\ref{vhfs}).
If necessessary, they can be calculated to relative order $\alpha (Z\alpha)^2$
suing the formalism recently outlined in Ref.~\cite{JeYe2010}.

The theoretical predictions for the transition frequencies in Rydberg states
are more precise than for lower-lying transitions, because a number of
problematic effects like the nuclear-size correction are effectively
suppressed.  It is thus possible to devise transitions in the infra-red (see
Tables~\ref{tableH}---\ref{tablem2}) for probing nuclear masses; these
transitions combine a favourable range of frequencies for the application of
optical frequency combs with small uncertainties in the theoretical QED
predictions. Summarizing the theoretical calculations, we can say 
that with an entirely realistic relative uncertainty of 
$1.4 \times 10^{-13}$ for the measurement of the transitions 
in deuterium and hydrogen, the electron to proton mass ratio 
could be determined with a relative uncertainty of 
$3.6 \times 10^{-10}$. This would mean a small improvement 
over the present value with a relative uncertainty of 
$4.3 \times 10^{-10}$. For the electron to deuteron mass ratio,
a relative uncertainty comparable to the 
2006 CODATA value could be achieved.
If it were possible to improve the accuracy further and to
reach a relative uncertainty of $2 \times 10^{-14}$ for 
the two-photon transition in hydrogen-like ${}^7$Li,
we could improve lithium nuclear masses
by  more than one order of magnitude 
as compared to their values in the AME2003 \cite{Audi2003337} 
adjustment and of similar accuracy as 
\cite{NaEtAl2006}. This would potentially confirm 
the large difference between these two values for the ${}^7$Li 
mass from Refs.~\cite{NaEtAl2006,Audi2003337} by an independent method.
As evident from the entry in the fifth row of Table~\ref{tablem2},
the accuracy could be further
improved with a more precise value for the electron mass
whose accurate determination is currently being pursued, e.g.,
in precise measurements of the bound-electron $g$ factors.
The mass determination of ${}^{10}$C using the isotope shift
would allow us to reach a relative uncertainty comparable 
to the current relative uncertainty of its mass 
of $4.0 \times 10^{-8}$ \cite{Audi2003337} if the transition 
frequency could be measured with a relative uncertainty 
of $1.4 \times 10^{-12}$.

As evident from the last two rows of Tables~\ref{tablem1},~\ref{tablem2},
~\ref{tablem3D} and~\ref{tablem3H}, one loses about four decimals 
in the determination of a nuclear 
mass from transition frequencies, due to inevitable numerical 
cancellations. For {\em method II}, this is explained in the text 
following Eq.~\eref{loss}. 
Nevertheless, the approach leads to very promising
results. With the progress made in recent years in reducing the relative
uncertainties in the frequency measurement using frequency combs 
(see \cite{HaNob2006} and references therein), there is justified hope for
even more precise results for transition frequencies in the future. In
turn, they would allow to further increase the accuracy of such a mass
determination. 
In view of the present efforts of measuring nuclear masses
with increased accuracy, it certainly would be helpful to have an
independent method to check the nuclear masses determined by comparing
cyclotron frequencies in Penning traps \cite{BrGa1986}.

Concerning the broader impact of the current proposal, we remember that
${}^7$Li is one of the primordial elements, and a more precise mass measurement
may help to improve our understanding of the processes in the Big Bang
Nucleosynthesis, as well as the reason for the low binding energy of the
discussed nucleus.  Moreover, ${}^7$Li also plays a role in many nuclear
reaction processes especially in the fusion cycle of the sun where an improved
mass measurement can help to find bounds for the mass of the solar
neutrino~\cite{Re1967} and also for the determination of other masses through
nuclear reactions~\cite{BaDaFoHa1972}.  The electron to proton and electron to
deuteron mass ratios are important for many applications in spectroscopy where
these two isotopes are studied very intensively. For example, a more accurate
electron to proton mass ratio might be very interesting for the planned
comparison of transition frequencies in hydrogen and anti-hydrogen
\cite{GaEtAl2008,BlKoRu1999} in order to test $CPT$ invariance. Both mass
ratios also play a role in the analysis of the hydrogen-deuterium isotope shift
which is used to deduce the difference of the mean square charge radii of the
proton and the deuteron~\cite{HuEtAl1998}.  Another field where precise nuclear
masses are of great interest is the study of the unitarity of the CKM
matrix~\cite{AbEtAl2004,Blaum2006}, where super-allowed beta decays are being
investigated to determine the CKM matrix element $V_{ud}$. The $^{10}$C nucleus
is of particular interest in these studies \cite{HaTo2006}. A further increase
in accuracy for the nuclear mass $m_N({}^{10}{\textrm{C}})$ can improve the
bounds on CKM non-unitarity and therefore provide a check for the standard
model.

In order to provide reliable theoretical predictions for such joint
experimental-theoretical efforts, we have performed analytic as well as
numerical calculations of the self-energy remainder function for states with
$n=5,\dots,8$ with $\ell=n-1$ and $\ell=n-2$, and we have also obtained values
for the $A_{60}$ coefficient for these states. The theoretical predictions for
the notoriously problematic self-energy remainder functions show an excellent
mutual agreement. This leaves little room for any conceivable changes in the
theoretical predictions due to calculational errors whose existence in complex QED calculations otherwise cannot be ruled out without extensive cross checks.  
The field appears to be open for experimental studies.

%
%

\section*{Acknowledgments}

Insightful discussions with P.~J.~Mohr and J.~N.~Tan are
gratefully acknowledged.
This research has been supported by the National Science 
Foundation (Grant PHY--~8555454) and by a Precision Measurement 
Grant from the National Institute of Standards and Technology.

 \section*{References}

\providecommand{\newblock}{}



\begin{thebibliography}{10}
\expandafter\ifx\csname url\endcsname\relax
  \def\url#1{{\tt #1}}\fi
\expandafter\ifx\csname urlprefix\endcsname\relax\def\urlprefix{URL }\fi
\providecommand{\eprint}[2][]{\url{#2}}

\bibitem{NiEtAl2000}
Niering M, Holzwarth R, Reichert J, Pokasov P, {Th~Udem}, Weitz M, H\"{a}nsch
  T~W, Lemonde P, Santarelli G, Abgrall M, Laurent P, Salomon C and Clairon A
  2000 {\em Phys. Rev. Lett.\/} {\bf 84} 5496--5499

\bibitem{WeEtAl1995}
Weitz M, Huber A, Schmidt-Kaler F, Leibfried D, Vassen W, Zimmermann C,
  Pachucki K, H\"ansch T~W, Julien L and Biraben F 1995 {\em Phys. Rev. A\/}
  {\bf 52} 2664--2681

\bibitem{HuEtAl1998}
Huber A, {Th~Udem}, Gross B, Reichert J, Kourogi M, Pachucki K, Weitz M and
  H\"{a}nsch T~W 1998 {\em Phys. Rev. Lett.\/} {\bf 80} 468--471

\bibitem{BeEtAl1997}
de~Beauvoir B, Nez F, Julien L, Cagnac B, Biraben F, Touahri D, Hilico L, Acef
  O, Clairon A and Zondy J~J 1997 {\em Phys. Rev. Lett.\/} {\bf 78} 440--443

\bibitem{ScEtAl1999}
Schwob C, Jozefowski L, de~Beauvoir B, Hilico L, Nez F, Julien L, Biraben F,
  Acef O, Clairon A and Zondy J~J 1999 {\em Phys. Rev. Lett.\/} {\bf 82}
  4960--4963

\bibitem{BoEtAl1996}
de~Beauvoir B, Nez F, Julien L, Cagnac B, Biraben F, Touahri D, Hilico L, Acef
  O, Clairon A and Zondy J~J 1997 {\em Phys. Rev. Lett.\/} {\bf 78} 440--443

\bibitem{BeHiBo1995}
Berkeland D~J, Hinds E~A and Boshier M~G 1995 {\em Phys. Rev. Lett.\/} {\bf 75}
  2470--2473

\bibitem{HaPi1994}
Hagley E~W and Pipkin F~M 1994 {\em Phys. Rev. Lett.\/} {\bf 72} 1172

\bibitem{LuPi1986}
Lundeen S~R and Pipkin F~M 1986 {\em Metrologia\/} {\bf 22} 9

\bibitem{NeAnUn1979}
Newton G, Andrews D~A and Unsworth P~J 1979 {\em Phil. Trans. R. Soc. A\/} {\bf
  290} 373

\bibitem{MoTaNe2008}
Mohr P~J, Taylor B~N and Newell D~B 2008 {\em Rev. Mod. Phys.\/} {\bf 80} 633

\bibitem{Pa1996PRA}
Pachucki K 1996 {\em Phys. Rev. A\/} {\bf 53} 2092

\bibitem{Pa1999PRA}
Pachucki K 1999 {\em Phys. Rev. A\/} {\bf 60} 3593

\bibitem{FrMaSp1997}
Friar J~L, Martorell J and Sprung D~W~L 1997 {\em Phys. Rev. A\/} {\bf 56}
  4579--4586

\bibitem{JeKoLBMoTa2005}
Jentschura U~D, Kotochigova S, Le~Bigot E~O, Mohr P~J and Taylor B~N 2005 {\em
  Phys. Rev. Lett.\/} {\bf 95} 163003

\bibitem{JeMoTaWu2008}
Jentschura U~D, Mohr P~J, Tan J~N and Wundt B~J 2008 {\em Phys. Rev. Lett.\/}
  {\bf 100} 160404

\bibitem{JeMoTaWu2009}
Jentschura U~D, Mohr P~J, Tan J~N and Wundt B~J 2009 {\em Can. J. Phys.\/} {\bf
  87} 757--762

\bibitem{GaEtAl2008}
Gabrielse G, Larochelle P, Le~Sage D, Levitt B, Kolthammer W~S, McConnell R,
  Richerme P, Wrubel J, Speck A, George M~C, Grzonka D, Oelert W, Sefzick T,
  Zhang Z, Carew A, Comeau D, Hessels E~A, Storry C~H, Weel M and Walz J 2008
  {\em Phys. Rev. Lett.\/} {\bf 100} 113001

\bibitem{Re1967}
Reines F 1967 {\em Proceedings of the Royal Society of London. Series A,
  Mathematical and Physical Sciences\/} {\bf 301} 159--170

\bibitem{NaEtAl2006}
Nagy S, Fritioff T, Suhonen M, Schuch R, Blaum K, Bj\"orkhage M and Bergstr\"om
  I 2006 {\em Phys. Rev. Lett.\/} {\bf 96} 163004

\bibitem{Audi2003337}
Audi G, Wapstra A~H and Thibault C 2003 {\em Nuclear Physics A\/} {\bf 729} 337
  -- 676

\bibitem{Blaum2006}
Blaum K 2006 {\em Physics Reports\/} {\bf 425} 1 -- 78

\bibitem{HaTo2006}
Hardy J~C and Towner I~S 2006 {\em Nucl. Phys. News\/} {\bf 16N4} 11--17

\bibitem{KoEtAl2007}
Koelemeij J~C~J, Roth B, Wicht A, Ernsting I and Schiller S 2007 {\em Phys.
  Rev. Lett.\/} {\bf 98} 173002

\bibitem{KaEtAl2008}
Karr J~P, Bielsa F, Douillet A, Pedregosa~Gutierrez J, Korobov V~I and Hilico L
  2008 {\em Phys. Rev. A\/} {\bf 77} 063410

\bibitem{FlEtAl2008}
Flowers J, Baird P, Klein H, Langham C, Margolis H and Walton B 2008 {\em
  Conference on Precision Electromagnetic Measurements Digest, 2008. CPEM
  2008.\/}  40--41

\bibitem{DKpc}
Kleppner D (private communicaton)

\bibitem{DVrPhD2001}
Vries J~C~D Ph.D. thesis, MIT, 2001

\bibitem{EiGrSh2001}
Eides M~I, Grotch H and Shelyuto V~A 2001 {\em Phys. Rep.\/} {\bf 342} 63--261

\bibitem{SaYe1990}
Sapirstein J and Yennie D~R 1990 Theory of hydrogenic bound states {\em Quantum
  Electrodynamics\/} ({\em Advanced Series on Directions in High Energy
  Physics\/} vol~7) ed Kinoshita T (Singapore: World Scientific) pp 560--672

\bibitem{BaGl1955}
Barker W~A and Glover F~N 1955 {\em Phys. Rev.\/} {\bf 99} 317

\bibitem{Er1977}
Erickson G~W 1977 {\em J. Phys. Chem. Ref. Data\/} {\bf 6} 831

\bibitem{PaGr1995}
Pachucki K and Grotch H 1995 {\em Phys. Rev. A\/} {\bf 51} 1854

\bibitem{GoElMiKh1995}
Golosov E, Elkhovskii A~S, Milshtein A~I and Khriplovich I~B 1995 {\em Zh.
  \'{E}ksp. Teor. Fiz.\/} {\bf 107} 393 [JETP {\bf 80} (2), 208 (1995)]

\bibitem{JePa1996}
Jentschura U and Pachucki K 1996 {\em Phys. Rev. A\/} {\bf 54} 1853--1861

\bibitem{WiKr1956}
Wichmann E~H and Kroll N~M 1956 {\em Phys. Rev.\/} {\bf 101} 843

\bibitem{Je2006}
Jentschura U~D 2006 {\em Phys. Rev. A\/} {\bf 74} 062517

\bibitem{WuJe2008}
Wundt B and Jentschura U~D 2008 {\em Phys. Lett. B\/} {\bf 659} 571--575

\bibitem{JeYe2006}
Jentschura U~D and Yerokhin V~A 2006 {\em Phys. Rev. A\/} {\bf 73} 062503

\bibitem{BrPa1967}
Brodsky S~J and Parsons R~G 1967 {\em Phys. Rev.\/} {\bf 163} 134--146

\bibitem{JeYe2010}
Jentschura U~D and Yerokhin V~A 2010 {\em Phys. Rev. A\/} {\bf 81} 012503

\bibitem{CaLe1986}
Caswell W~E and Lepage G~P 1986 {\em Phys. Lett. B\/} {\bf 167} 437

\bibitem{JeCzPa2005}
Jentschura U~D, Czarnecki A and Pachucki K 2005 {\em Phys. Rev. A\/} {\bf 72}
  062102

\bibitem{SaOe1989}
Salomonson S and \"{O}ster P 1989 {\em Phys. Rev. A\/} {\bf 40} 5559

\bibitem{JeMoTa2009}
Jentschura U~D, Mohr P~J and Tan J~N 2009 {\em J. Phys. B\/}  in press

\bibitem{HaFoGa2008}
Hannecke D, Fogwell S and Gabrielse G 2008 {\em Phys. Rev. Lett.\/} {\bf 100}
  120801

\bibitem{MoTa2005}
Mohr P~J and Taylor B~N 2005 {\em Rev. Mod. Phys.\/} {\bf 77} 1

\bibitem{CRC}
Lide D~R 2007 {\em CRC Handbook of Chemistry and Physics, 88th Edition (Crc
  Handbook of Chemistry and Physics)\/} (Boca Raton, FL: CRC)

\bibitem{FiEtAl2004}
Fischer M, Kolachevsky N, Zimmermann M, Holzwarth R, {Th~Udem}, H\"{a}nsch T~W,
  Abgrall M, Gr\"unert J, Maksimovic I, Bize S, Marion H, Pereira Dos~Santos F,
  Lemonde P, Santarelli G, Laurent P, Clairon A, Salomon C, Haas M, Jentschura
  U~D and Keitel C~H 2004 {\em Phys. Rev. Lett.\/} {\bf 92} 230802

\bibitem{VDyEtAl2006}
Dyck R~S~V, Jr, Pinegar D~B, Liew S~V and Zafonte S~L 2006 {\em International
  Journal of Mass Spectrometry\/} {\bf 251} 231 -- 242

\bibitem{VeEtAl2004}
Verd\'{u} J, Djeki\'{c} S, Stahl S, Valenzuela T, Vogel M, Werth G, Beier T,
  Kluge H~J and Quint W 2004 {\em Phys. Rev. Lett.\/} {\bf 92} 093002

\bibitem{VoEtAl2005}
Vogel M, Alonso J, Djekic S, Kluge H~J, Quint W, Stahl S, Verdu J and Werth G
  2005 {\em Nuclear Instruments and Methods in Physics Research Section B: Beam
  Interactions with Materials and Atoms\/} {\bf 235} 7 -- 16

\bibitem{WeEtAl2006}
Werth G, Alonso J, Beier T, Blaum K, Djekic S, Häffner H, Hermanspahn N, Quint
  W, Stahl S, Verdú J, Valenzuela T and Vogel M 2006 {\em International Journal
  of Mass Spectrometry\/} {\bf 251} 152 -- 158

\bibitem{QuNiJe2008}
Quint W, Nikoobakht B and Jentschura U~D 2008 {\em Pis'ma v. Zh. \'{E}ksp.
  Teor. Fiz.\/} {\bf 87} 36--40 [JETP Lett. {\bf 87}, 30 (2008)]

\bibitem{QuEtAl2008}
Quint W, Moskovkhin D~L, Shabaev V~M and Vogel M 2008 {\em Phys. Rev. A\/} {\bf
  78} 032517

\bibitem{JeEtAl2005}
Jentschura U~D, Le~Bigot E~O, Evers J, Mohr P~J and Keitel C~H 2005 {\em J.
  Phys. B\/} {\bf 38} S97--S105

\bibitem{BeSa1957}
Bethe H~A and Salpeter E~E 1957 {\em Quantum Mechanics of One- and Two-Electron
  Atoms\/} (Berlin: Springer)

\bibitem{LuPi1981}
Lundeen S~R and Pipkin F~M 1981 {\em Phys. Rev. Lett.\/} {\bf 46} 232

\bibitem{HaNob2006}
Haensch T~W 2006 {\em Rev. Mod. Phys.\/} {\bf 78} 1297

\bibitem{BrGa1986}
Brown L~S and Gabrielse G 1986 {\em Rev. Mod. Phys.\/} {\bf 58} 233--311

\bibitem{BaDaFoHa1972}
Ball G~C, Davies W~G, Forster J~S and Hardy J~C 1972 {\em Phys. Rev. Lett.\/}
  {\bf 28} 1069--1071

\bibitem{BlKoRu1999}
Bluhm R, Kosteleck\'y V~A and Russell N 1999 {\em Phys. Rev. Lett.\/} {\bf 82}
  2254--2257

\bibitem{AbEtAl2004}
Abele H {\em et~al.\/} 2004 {\em Eur. Phys. J.\/} {\bf C33} 1--8
  (\textit{Preprint} \eprint{hep-ph/0312150})

\end{thebibliography}
\end{document}